\documentclass{webofc}

\usepackage[varg]{txfonts}   
\usepackage{hyperref}
\usepackage{url}
\hypersetup{colorlinks=true,citecolor=blue,urlcolor=blue,linkcolor=blue}
\begin{document}
\title{Non-Equilibrium Aspects of Fission Dynamics within the Time Dependent Density Functional Theory}

\author{ \firstname{Aurel} \lastname{Bulgac} \inst{1}  \fnsep\thanks{\email{bulgac@uw.edu} }  
 \and
               \firstname{Matthew} \lastname{Kafker}\inst{1}                         \and
               \firstname{Ibrahim} \lastname{Abdurrahman}\inst{2}            
               \and
               \firstname{Ionel} \lastname{Stetcu}\inst{2}                      }

\institute{ Department of Physics, University of Washington, Seattle, WA 98195, USA
\and 
Theoretical Division, Los Alamos National Laboratory, Los Alamos, NM 87545, USA
}

\abstract{ We will cover briefly the time-dependent density functional theory approach, extended to include pairing correlations, 
to induced fission of $^{238}$U(n,f), $^{241,243}$Pu, and $^{238}$Np, and also results on scission neutrons and a number 
of very nontrivial aspects of induced fission.  

}

%
\maketitle
\section{Introduction}
\label{intro}

For all results reported here we performed extensive fully microscopic studies of fission dynamics within the time-dependent 
density functional theory (TDDFT) extended to include pairing correlations, with controlled numerical approximations and without 
introducing any unchecked assumptions or simplifications. The theoretical framework is described and was reviewed in great 
detail in Refs.~\cite{Bulgac:2019,Shi:2020,Bulgac:2018}  and the corresponding codes are accessible on GitHub.

\section{Induced fission of odd-mass and odd-odd nuclei}

We performed  studies of the induced fission dynamics  of nuclei with either odd number of neutrons  
for the compound fission systems $^{238}$U(n,f) [23 trajectories], $^{240}$Pu(n,f) [42 trajectories], 
$^{242}$Pu(n,f) [7 trajectories], and with both odd proton  and neutron numbers for 
$^{237}$Np(n,f) [48 trajectories], where in each case the number of trajectories describe various initial  nuclear shape deformations 
characterized by quadrupole and octupole momenta $(Q_{20}, Q_{30})$ near the outer fission barrier
and various odd quasiparticle states of the respective fissioning compound nucleus, constructed using the standard prescription 
outlined in Refs.~\cite{Ring:2004, Bertsch:2009,Schunck:2023,Pore:2024}.  In addition, for comparison, 
 we have performed additional calculations also for  the induced fission of
$^{235}$U(n,f) [24 trajectories], $^{239}$Pu(n,f)  [7 trajectories] and for the odd nucleon systems $^{237}$Np(n,f) [11 trajectories], 
$^{238}$U(n,f) [6 trajectories], $^{240}$Pu(n,f) [13 trajectories], and $^{242}$Pu(n,f) [2 trajectories], treating them as even-even nuclei for a total of 63 trajectories.  
The total 121+63 = 184 TDDFT trajectories for these nuclei is comparable (or maybe even bigger) with perhaps of all 
of the rest of the world TDDFT fission trajectories obtained for even-even nuclei by other authors since 2015, see review talk  of D. Vretenar for some references. 
The simulations have been performed in a simulation box $30^2\times60$ fm$^3$ with a lattice constant of 1 fm 
and following the dynamics for up to about 30,000 to 60,000 times steps or longer in some cases. These 
simulations were performed with the energy density functional SeaLL1~\cite{Bulgac:2018} by evolving $2\times 2 \times 4\times 30^2\times 60=864,000$ complex, coupled, 
nonlinear PDEs in 3D+time, see Refs~\cite{Bulgac:2018,Bulgac:2019,Shi:2020,Bulgac:2019c,Bulgac:2020,Bulgac:2023,Bulgac:2024a} for details. 
This proves that inspite of the complexity of these 
simulations, using a total number of quasiparticle states $2\times 2\times 30^2\times 60 =216,000$, such simulations can be performed 
on a large variety of supercomputers available worldwide nowadays.
Detailed extensive results of these simulations will be submitted for publication soon.

\begin{figure*}[h]
\centering
\vspace*{1cm}       
\includegraphics[width=6.25cm,clip]{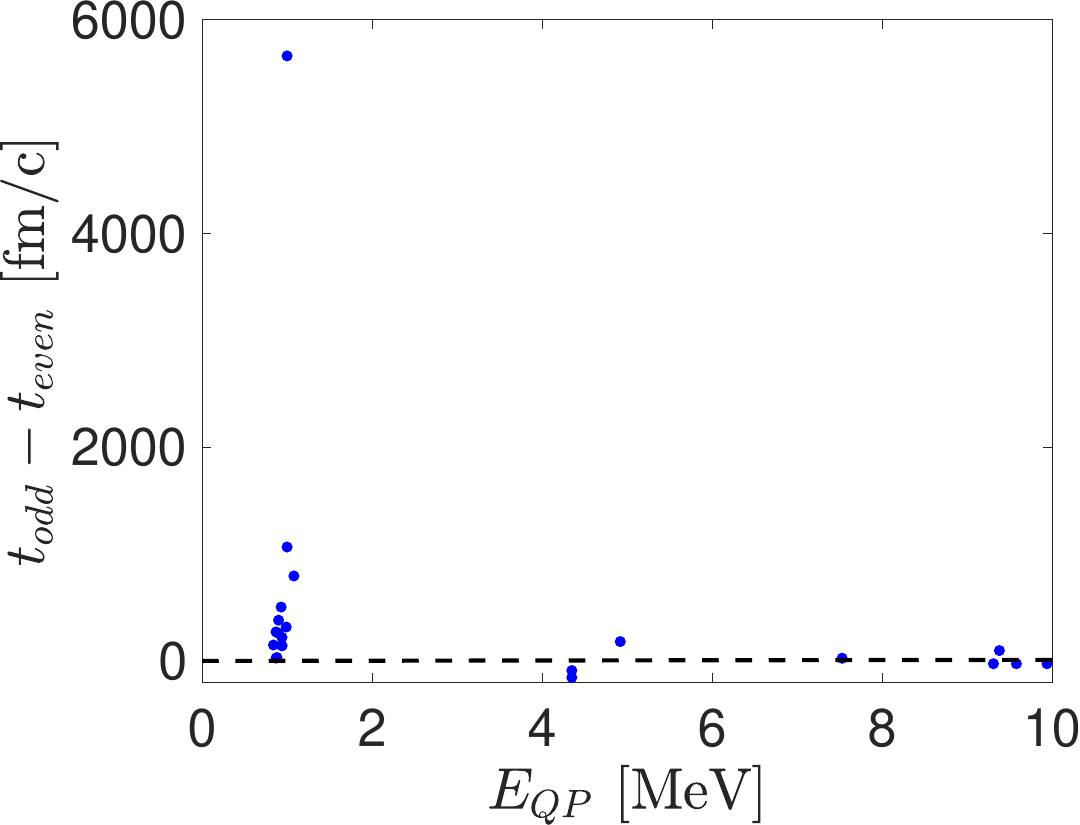}
\includegraphics[width=6.25cm,clip]{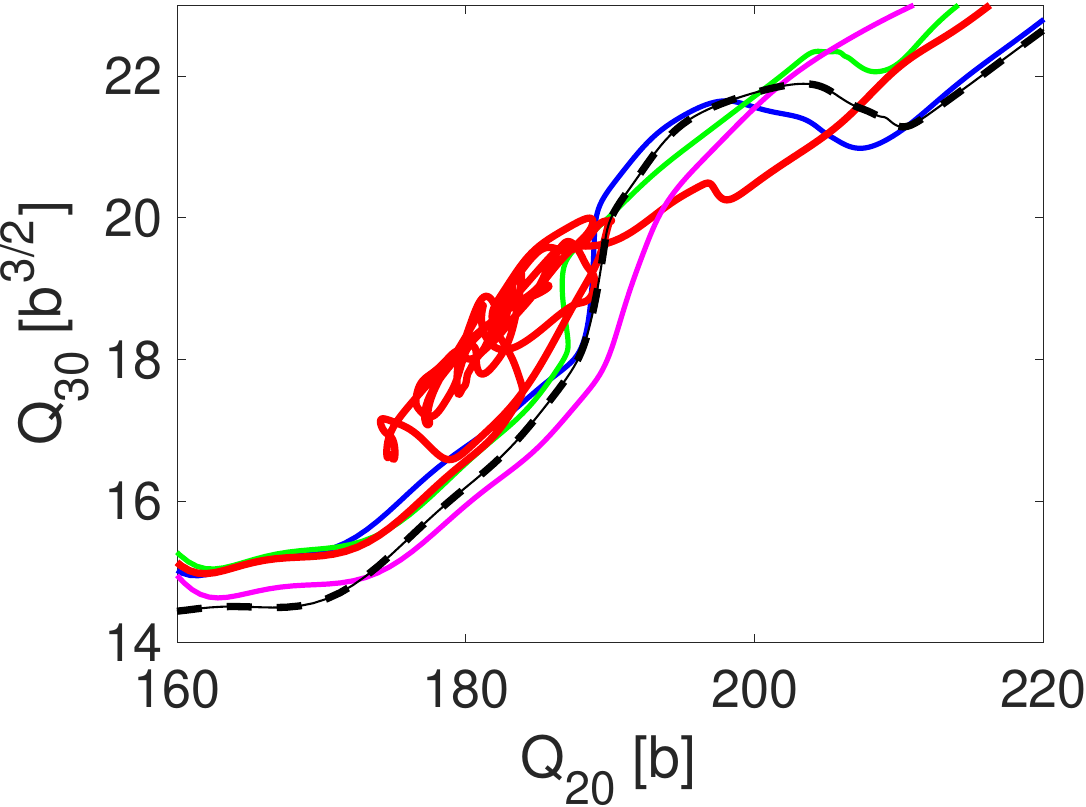}
\includegraphics[width=7.5cm,clip]{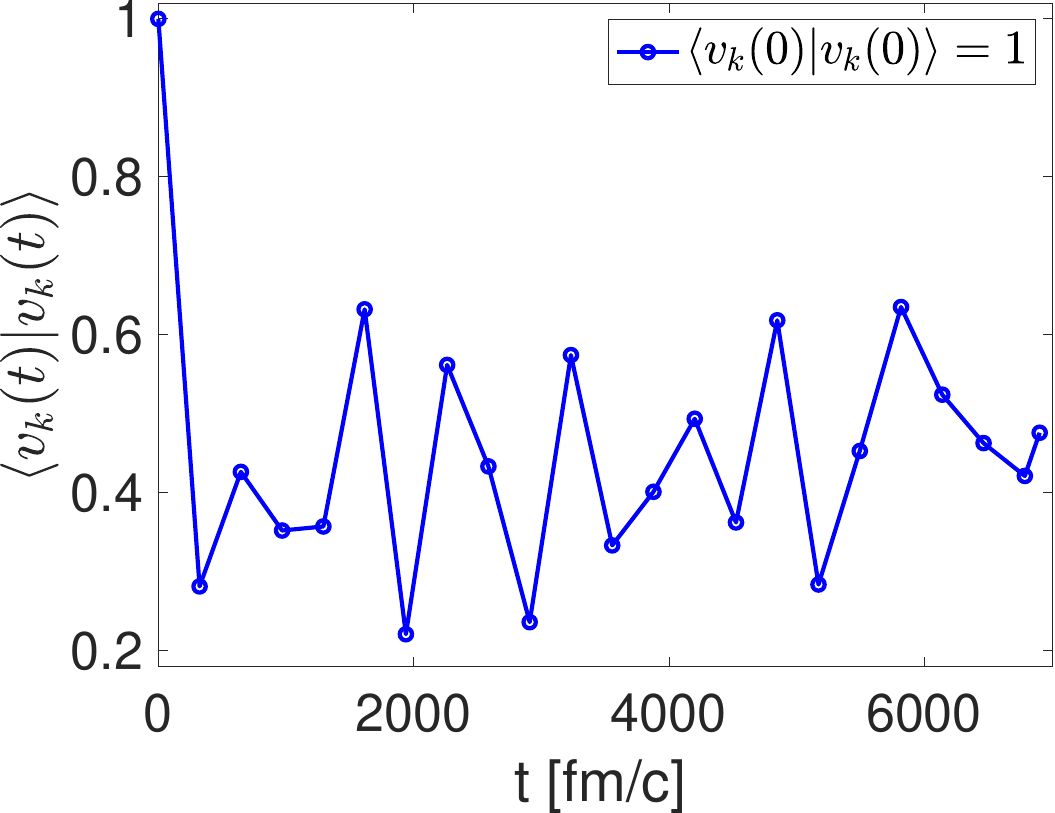}
\caption{
In the upper row left panel we display the time difference to full spatial separation of the two fission fragments (FFs)
 between the case of an even-even and odd-mass nuclei. 
The odd-mass system is treated with an exact odd-neutron and even-proton numbers respectively.
The odd-mass nucleus is treated as having an even-neutron and an even-proton  numbers, 
with the average $N=147$ and $Z=92$ respectively for $^{239}$U and then a quasiparticle state is excited as described in Refs.~\cite{Ring:2004,Bertsch:2009}.
In the upper row right panel  we show several TDDFT trajectories for the fission of an odd-nucleon nucleus with 
solid lines and for reference a TDDFT trajectory for an even-even nucleus with average $N=147$ and $Z=92$, with a dashed-thin solid line. 
Unlike the TDDFT trajectories, in which typically the quadrupole moment $Q_{20}$ typically increases in time in 
the case of fission of odd-nucleon systems the nucleus encounters a ``bumpier'' potential energy surface in the $(Q_{20},Q_{30})$ 
plane towards scission and the system it gets ``confused'' following a much more "challenging obstacle course," see upper row right panel.
In the second row we show the occupation probability of the initial  odd-nucleon state as a function of time. For time $t=0$ the occupation probability was obtained by 
constructing the corresponding canonical occupation probabilities~\cite{Bulgac:2023}. For $t>0$ apart from one canonical quasiparticle state with occupation 
probability 1, the rest of the quasiparticle states are all double  occupied with probabilities less than 1. }
\label{fig-one}       
\end{figure*}

Ever since the development of a microscopic description of pairing correlations by Bardeen {\it et al.}~\cite{Bardeen:1957} 
and the recognition that they are also most likely crucial 
in explaining nuclear properties~\cite{Bohr:1958}, the treatments of fermion systems with odd numbers of particles have relied on the so called Pauli blocking approximation. 
The extra odd nucleon carries a non-vanishing angular momentum and as a result the nuclear mean field breaks time-reversal invariance. 
The compound nucleus breaks the gauge symmetry as well and has a negative particle parity. The 
rotational  symmetry and sometimes spatial parity are also broken. From the theoretical point of view 
these are rather inconvenient features, resulting in a significant number of technical difficulties, and 
which forced most theoretical treatments to adopt approximations and assumptions.  
In nuclear large amplitude collective motion the Pauli blocking approximation has been used almost always, see the most recent and detailed 
studies of the fission of odd-mass nuclei, where Pauli blocking in conjunction along with the 
equal filling approximation~\cite{Perez-Martin:2008} have been 
implemented~\cite{Schunck:2023,Pore:2024}.  

Overall the FF properties in the induced fission of even-even, odd-mass, and odd-odd nuclei are very similar, 
but the widths of total kinetic energy, FF neutron and charge distributions are typically twice as large as in the case of induced fission of even-even nuclei. 
Often there is a difference in the times the two FFs reach scission, when the corresponding initial quasiparticle state 
has a relatively low excitation energy, typically smaller than the pairing gap, see Fig. \ref{fig-one}.  
In that case the TDDFT trajectory is significantly more complicated as the system spends quite a lot of time 
``deciding''' which path to follow on the potential energy surface. Such a situation 
has been encountered for some almost symmetric TDDFT trajectories~\cite{Bulgac:2019c,Bulgac:2020} 
in the case of fission even-even compound nuclei. 

An unexpected feature is the violation of the Pauli blocking approximation, see Fig. \ref{fig-one}, used for more than 6 decades in nuclear physics. During the 
time evolution, as was  demonstrated in Ref.~\cite{Bulgac:2024a}, even if the system is started in a fully canonical basis, in the subsequent time evolution the quasiparticle  
states are not anymore canonical states, and at each time step one has to identify the new canonical set of wave functions. One can clearly see that the 
single odd quasiparticle state initially occupied with probability 1 in canonical basis, 
ceases to maintain the same occupation probability. However, at each time $t>00$ there is only one canonical state 
occupied with exactly probability 1. We expect that this is a feature to be observed in any nuclear Large Amplitude Collective Motion (LACM) of an odd fermion system, 
in either excited states of a nucleus or in heavy-ion collisions. A clear example when such a situation is encountered is when the angular momentum of the 
initial odd fermion is $\approx A^{1/3}$, since such a state cannot exist with a significant occupation probability in either 
FF~\cite{Bertsch:1997,Bulgac:2016,Bulgac:2019c,Bulgac:2020}.

\section{Scission neutrons and nontrivial aspects of fission dynamics revealed in TDDFT framework}
 The scission neutrons have been conjectured to exist by Bohr and Wheeler~\cite{Bohr:1939} and more than 50 years after discovery of nuclear fission
 Wagemans~\cite{Wagemans:1991} stated (see his introductory remarks) that they most likely do not exist. 
Wagemans's remarks are notable in another respect also, he never refers to Meitner and Frisch breakthrough paper~\cite{Meitner:1939} and instead gives all the credit to 
Bohr and Wheeler~\cite{Bohr:1939} for explaining the nature of nuclear fission, even though Bohr and Wheeler refer to Meitner and Frisch in their paper
and even the name nuclear fission was coined by Meitner and Frisch. 
The experimental search and various theoretical models predicted the existence of scission neutrons with contradicting conclusions and properties in the following years, 
as either dominating the number of emitted neutrons by fission fragments (FFs) or being almost absent. In a very recent paper~\cite{Abdurrahman:2024} we have reported 
the results of our most comprehensive theoretical analysis of scission neutrons within the Time-Dependent Density Functional Theory (TDDFT) extended to superfluid 
fermionic systems and predicted an extensive range of their properties. We invite the reader to consult this publication~\cite{Abdurrahman:2024}, 
including the quite extensive supplement.

In recent publications~\cite{Bulgac:2023,Bulgac:2024a} we have presented several new aspects of the fission process, which were never discussed in literature. 
We have demonstrated that the fission dynamics from saddle-to-scission has an intrinsically non-Markovian 
character and consequently various previous attempts to describe this process using various stochastic or kinetic approaches are inadequate. 
Moreover, since in nuclear LACM the occupation probabilities of single-particle states with large momenta have very long  tails $\propto  C/p^4$, where $p$ is the local 
single-particle momentum, a significant repopulation of the quasiparticle states occurs during the fission 
dynamics and theoretical modeling with a reduced basis set of single-particle states, 
such as (TD)BCS+TDHF or even TDDFT with a limited number of quasiparticle states 
lead to inaccurate predictions of the FFs properties, or even fail to lead to 
fission. 

\section{Conclusions}
TDDFT, extended to include a correct description of the dynamics of the pairing correlations, is arguably 
the most advanced microscopic quantum tool available today, and it is based on solid independent knowledge of  
basic nuclear properties of nuclei with $A\geq 16$.  In this approach 
only 7 parameters well known  for a long time  are needed: 
energy and equilibrium density of symmetric nuclear matter, surface tension, 
symmetry energy and its density dependence, spin-orbit and pairing strengths), and this leads to a one of the most accurate descriptions 
of the  binding energies, charge radii, one- and two-neutron separation energies, single-particle spectra of 
magic nuclei, etc.,  see Ref.~\cite{Bulgac:2018}, and it is used in TDDFT  to describe fission dynamics. 
We described briefly several recent developments in a fully microscopic description of fission dynamics. 
i)  The TDDFT approach to the emission of scission neutrons, a topic still not settled  either theoretically 
or experimentally, was conjectured in 
1939~\cite{Bohr:1939} soon after nuclear fission explanation of fission~\cite{Meitner:1939}.
 ii) The non-Markovian aspects of the fission descent, 
which is distinct from the earlier demonstrated  clear dissipative character of fission dynamics. 
iii) A first peak to some of our preliminary results from the first TDDFT treatment 
of the induced fission of a large class of odd nucleon compound nuclei.  
A overall arching conclusion emerges from these studies: fission dynamics is an 
intrinsically non-equilibrium process, which requires a detailed time-dependent microscopic description,
without the inclusion of unchecked assumptions, phenomenology and parameter fitting, and 
uncontrolled approximations and simplifications.\\
 
The funding for A.B. and M.K. from the Office of Science, Grant No. DE-FG02-97ER41014  
and also the partial support provided by NNSA cooperative Agreement DE-NA0003841 is greatly appreciated. 
 The work of I.A. and I.S. was supported by the U.S.
Department of Energy through the Los Alamos National
Laboratory. The Los Alamos National Laboratory is operated
by Triad National Security, LLC, for the National Nuclear
Security Administration of the U.S. Department of
Energy Contract No. 89233218CNA000001.
This research used resources of the Oak Ridge Leadership Computing Facility, which is a U.S. DOE Office of
Science User Facility supported under Contract No. DE-AC05-00OR22725.

\providecommand{\selectlanguage}[1]{}
\renewcommand{\selectlanguage}[1]{}

\bibliography{local_fission}
 
\end{document}